\documentclass[pdflatex,sn-standardnature]{sn-jnl}

\usepackage{subfigure}
\DeclareMathOperator{\sgn}{sgn}

\jyear{2022}

\begin{document}

\title[Implementation and Applications of a Ternary Threshold Logic Gate]{Implementation and Applications of a Ternary Threshold Logic Gate}

\author*{\fnm{Ahmet} \sur{Unutulmaz}}\email{ahmet.unutulmaz@marmara.edu.tr}
\author{\fnm{Cem} \sur{\"{U}nsalan}}\email{cem.unsalan@marmara.edu.tr}

\affil{\orgdiv{Faculty of Engineering, Department of Electrical and Electronics Engineering}, \orgname{Marmara University}, \orgaddress{\city{Istanbul}, \country{Turkey}}}

\abstract{Reducing delay, power consumption, and chip area of a logic circuit are the main targets of a designer. Most of the times, the designer sacrifices power consumption and chip area to improve delay for a given technology node. To overcome this problem, we propose a ternary threshold logic gate. We implement the proposed gate by combining threshold logic and ternary logic. Then, we construct basic building blocks of a ternary ALU (as logic gates, comparator, and arithmetic circuits) using the proposed gate. We show that the proposed ternary TLG improves delay, power consumption, and chip area of ternary circuits via simulations. Thus, the proposed gate can be used to improve delay, power consumption, and chip area of ternary circuits.}


\maketitle

\section{Introduction}\label{sec:intro}

Decreasing delay of a logic circuit is one of the main targets of a designer. Most of the times, the designer sacrifices power consumption to improve delay for a given technology node. One way to overcome this trade-off is the utilization of binary threshold logic gates (TLG). Previous studies showed that delay and power consumption of binary logic circuits can be improved by utilizing TLG \cite{Leshner2010, Kulkarni2016}. We provide a brief review of threshold logic circuits in Section~\ref{sec:revTLG}. The designer should also keep track of the design cost in addition to improving delay and power consumption. The cost may be reduced by using less chip area. One way of doing this is via reducing wire (routing) congestion \cite{Saxena2007}. Ternary logic helps in this respect. A wire carrying a ternary signal (trit) can have three different logic levels, whereas a wire carrying a binary signal (bit) can only have two different logic levels. Therefore, information is spatially compressed in the ternary case. Thus, a ternary logic circuit requires less routing resources and chip area.

Researchers implemented ternary logic circuits via MOSFET and carbon nano-tube FET (CNTFET) technologies. Ternary circuit implementations with complementary MOSFETs (CMOS) dates backs to 1980s. Then, the available processes were only offering depletion and enhancement type\footnote{A depletion mode n-type device is on when gate to source voltage $V_{gs}$ is 0~V where an enhancement mode n-type device is off when $V_{gs}$ = 0~V.} devices, where ternary circuit implementations were requiring negative potential \cite{Balla1984}. For several decades, not much progress has been made in ternary logic implementations. We believe that this was mainly due to the limitations of older implementation technologies. Research in ternary logic gained momentum with the development of CNTFETs \cite{Lin2009, Gadgil2020}. However, practicality of large-scale CNTFET circuits is hard to achieve despite its inherent support for the multi threshold design paradigm. To the authors knowledge, there are no CNTFET implementations of large scale ternary logic circuits. In terms of binary logic, the largest fabricated CNTFET circuit contains only about 15.000 CNTFETs \cite{Hills2019}. This is far behind today's binary designs which utilizes modern MOSFET/FinFET technologies. Although, MOSFET technologies of 1980s were not very suitable for ternary logic circuits, recent MOSFET/FinFET technologies offer devices with different thresholds, even for 7~nm node and beyond \cite{Chang2020}. These multi-Vt devices provide an opportunity to decrease power consumption. They may also be used to implement ternary logic circuits. We provide a short overview of ternary logic implementations in Section~\ref{sec:revTernary}.

In this study, we aim to decrease delay, power consumption, and chip area of CMOS ternary circuits. We achieve these by combining advantages of threshold and ternary logic. Therefore, we propose a ternary threshold logic gate (TLG). The proposed ternary TLG is based on modern multi-Vt MOSFET technology. To our knowledge, this is the first study implementing ternary TLG using multi-Vt MOSFETs. We present the ternary TLG in Section~\ref{sec:ternaryTLG}. We then use it to implement basic blocks of an ALU. Therefore, we first implement the ternary inverter (STI), AND, OR, and XOR gates in Section~\ref{sec:logicTernary}. Next, we implement a ternary comparator in Section~\ref{sec:compTernary}. Then, we present ternary arithmetic circuits in Section~\ref{sec:arthTernary}. In Section~\ref{sec:comparison}, we compare the proposed ternary TLG based circuits with their CMOS only implementations and the ones in literature. Comparison results indicate that the proposed TLG based STI, XOR, comparator, and arithmetic circuits have significant advantages over their CMOS only implementations. Finally, we report key findings and future research directions in Section~\ref{section:final_comments}.

\section{Background}

The proposed ternary TLG is based on two main concepts: threshold logic and ternary logic. We summarize the previous work on these two concepts in this section. We start with the threshold logic and continue with ternary logic.

\subsection{Review of Threshold Logic}\label{sec:revTLG}

History of threshold logic and detailed review of its implementations are presented in \cite{Beiu2003}. There, implementations are categorized into five groups as CMOS solutions, capacitive implementations, conductance/current implementations, single electron tunneling solutions, and resonant tunneling devices. Differential solutions, a sub class of conductance/current implementations, turns out to be the most promising ones in terms of delay and power consumption. These are also known as differential threshold logic gates and abbreviated as DTL gates, differential TLG or simply TLG.

Differential TLG is suitable to be used with standard CMOS logic. Studies have shown that using threshold logic with standard CMOS gates reduces power consumption and necessary chip area \cite{Leshner2010, Kulkarni2016}. In \cite{Leshner2010}, a 32-bit two’s complement integer multiplier was designed using TLG in conjunction with CMOS standard cells on a 65~nm process node. Compared to CMOS only implementation, the design was 1.23 times smaller, consumed 1.41 times less dynamic power, and had 2.5 times less leakage power. The authors also fabricated their design. In a recent study \cite{Kulkarni2016}, a standard cell library including TLG was built. Here, an SR latch is abutted after the TLG. The library is then used to synthesize a Wallace tree multiplier, FIR filter, floating point multiplier, 32-bit MIPS core, and AES implementation. Synthesized circuits, including both CMOS and TLG cells, outperformed CMOS only implementations in terms of chip area usage, dynamic and leakage power consumption. A review of synthesis methods targeting threshold logic, as well as a logic synthesis flow which maps a digital circuit function to a threshold logic network, is presented in \cite{Neutzling2018}. The optimization method used in this study was able to estimate the chip area using different cost functions.

Popularity of threshold logic increased with emerging technologies such as memristors \cite{Mazumder2012}. A comprehensive survey on memristive threshold logic circuits was presented in \cite{Maan2016}. In \cite{Yang2014}, a TLG implementation is demonstrated. The authors claimed that utilization of a memristor network via TLG implementation improves robustness. Similarly, memristors were integrated into TLGs in \cite{Vrudhula2015} to increase the robustness of logic gates and improve power consumption. Recently, physical implementation of a memristor-based TLG is introduced in \cite{Papandroulidakis2019}. Here, in-house fabricated metal-oxide memristors were used in implementation. The authors demonstrated two, three, and four input TLGs.

\subsection{Review of Ternary Logic}\label{sec:revTernary}

First attempts to use multi-valued logic, specifically ternary logic, in computation systems have started several decades ago. The authors \cite{Balla1984} stated the shortcomings of previous (T-gate) approaches as their exponential complexity with the number of inputs. They utilized CMOS technology to implement ternary inverters, NAND, and NOR gates using enhancement and depletion mode transistors together. Then, they implemented a full adder using conventional binary logic with ternary-to-binary decoder/encoder circuits before/after the binary circuit. There, calculations were done in binary logic. Input/output signals were encoded in ternary logic. This improved connectivity and reduced number of input and output pins.

Multi-valued logic systems, specifically ternary logic, gained popularity with the development of CNTFETs. This is mainly due to the fact that threshold voltage of a CNTFET can be adjusted by changing its physical dimensions (more precisely, the chirality vector). Transistors with different threshold values will turn on/off at different voltage levels. This significantly eases implementation of multi-valued logic circuits. CNTFETs were used to design ternary inverter, NAND, and NOR gates in \cite{Lin2009}. These gates were then combined with binary gates to implement a half adder and one-bit multiplier. CNTFET based half and full-adders and a 3-trit multiplier are presented in \cite{Bastani2017}. The authors aimed to reduce the number of CNTFETs in \cite{Srinivasu2016}. They applied their method on a half and full adder circuit in addition to a 3-trit multiplier. A decoder based high-performance half adder was presented in \cite{Sahoo2017}. Memristor based two-bit ternary adder and multiplier circuits were implemented using CNTFETs in \cite{Soliman2018, Soliman2019}. In these studies, ternary outputs are obtained via memristor based gates. Voltage at the output of these gates were digitized using ternary inverters. The proposed methodology yielded compact circuits. However, inherent static power dissipation of memristor gates may limit their application. A ternary ALU design was presented in \cite{Gadgil2020}. Here, the authors made use of 2-to-1 multiplexers to convert binary and ternary logic instead of using a decoder/encoder based design. The multiplexer-based design methodology improved power consumption compared to encoder/decoder based designs in simulations. Multiplexer based adder and subtractor circuits were implemented in \cite{Sharma2019}. A ternary logic synthesizer targeting emerging device technologies was presented in \cite{Srinivasu2017}. The proposed method was based on a geometric (cubic) representation. The proposed synthesizer runs in two steps. First, the ternary function is expressed in terms of unary functions. Then, the implementation is optimized considering transistor count. Even though not based on ternary logic, a worth mentioning study is \cite{Hills2019}. The authors fabricated a binary RISC-V microprocessor with CNTFETs. To our knowledge, this is the largest CNTFET design containing approximately 15.000 CNTFETs. This study shows that CNTFET technology is promising. However, it also indicates that CNTFET technology is far behind advanced MOSFET/FinFET technology nodes which yield designs with several billions of transistors on a single die.

\section{Ternary Threshold Logic Gate}\label{sec:ternaryTLG}

Several studies have shown that replacing compatible parts of binary logic with TLGs reduces delay \cite{Beiu2003, Leshner2010, Kulkarni2016}. This is done without sacrificing the power consumption. Or the power consumption is reduced this way without increasing the delay. We extend this idea from binary to ternary logic in this section. Hence, we propose the ternary TLG. To our knowledge, this is the first study implementing ternary TLG using a multi-Vt CMOS process. This gate can be used to decrease delay and power consumption of ternary logic circuits. In this section, we first provide functional form of the proposed ternary TLG with its block diagram. Then, we provide its circuit implementation.

\subsection{Ternary TLG Function}\label{sec:blockTLG}

\begin{figure}[htbp]
	\centering
	\subfigure[Block diagram]{\includegraphics[width=0.6\columnwidth]{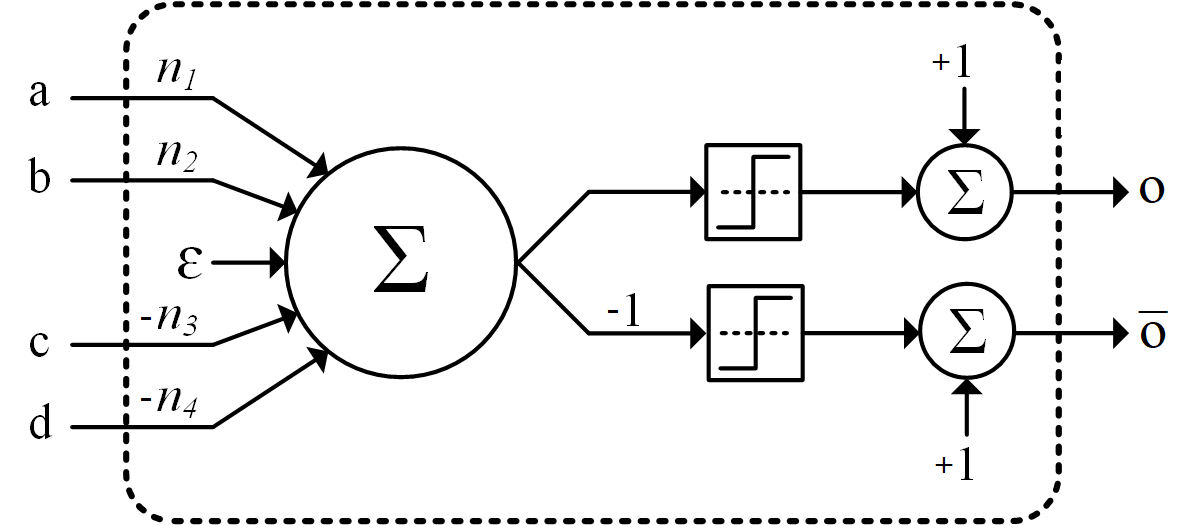}\label{fig:DTLBlock}}
	\subfigure[Symbol]{\includegraphics[width=0.2\columnwidth]{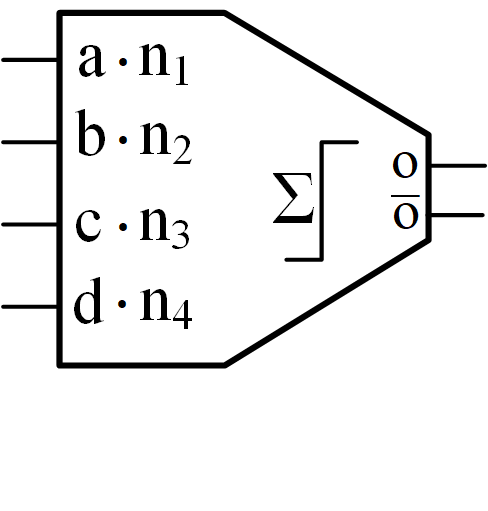}\label{fig:DTLSymbol}}
	\caption{Block diagram and symbol of the proposed ternary TLG.}\label{fig:DTL}
\end{figure}

Block diagram of the proposed ternary TLG is given in Fig.~\ref{fig:DTLBlock}. Inputs ($a$, $b$, $c$, $d$) are in ternary logic in this diagram. Outputs $o$ and $\bar{o}$ are in binary form and they are complement of each other. Note that, the binary value 0 (false) corresponds to ternary value 0. Whereas, the binary value 1 (true) corresponds to ternary value 2. Thus, outputs $o$ and $\bar{o}$ can take ternary values 0 or 2.

We can formulate the output $o$ as function of its inputs as

\begin{equation}\label{eqn:o_and_obar1}
o = \sgn(n_1 \cdot a + n_2 \cdot b + \varepsilon - n_3 \cdot c - n_4 \cdot d) + 1 
\end{equation}

\noindent where $\varepsilon$ is a small positive offset, $n_i$s are integer coefficients, and $\sgn(\cdot)$ is the sign function defined as

\begin{equation}
\sgn(x) =
\begin{cases}
   1 & \text{if } x > 0 \\
   0 & \text{if } x = 0 \\
  -1 & \text{if } x < 0
\end{cases}
\end{equation}

\noindent We will refer the proposed ternary TLG via the symbol in Fig.~\ref{fig:DTLSymbol} in the following sections.

\subsection{Circuit Implementation}\label{sec:circuitTLG}

\begin{figure}[htbp]
	\centering
	\includegraphics[width=1\columnwidth]{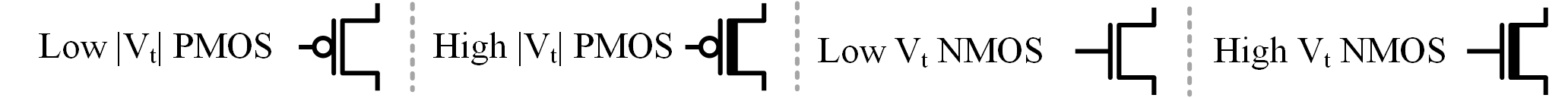}
	\caption{Symbols of low and high $V_t$ transistors.}\label{fig:tranSymb}
\end{figure}

Before dealing with circuit implementation, we provide details of the utilized technology. The proposed implementations are done using extended versions of ptm-32nm technology models \cite{Zhao2006, PTM}. Low threshold NMOS and PMOS devices are defined in addition to the original NMOS and PMOS models. Threshold voltages of these new devices are defined as $\lvert V_t \rvert$ = 0.25~V. Symbols for the original (high threshold) and new (low threshold) transistors are given in Fig.~\ref{fig:tranSymb}. In this study, voltage values corresponding to ternary logic levels 2, 1, and 0 are $V_{dd}$ = 1~V, $V_{bb}$ = 0.5~V and 0~V (ground), respectively.

We implement the proposed ternary TLG in Fig.~\ref{fig:DTLBlock} by combining ternary inverters with a differential TLG circuit. To feed the ternary signal to TLG, we decompose ternary signals into binary form. As an example, the ternary signal $a$ can be expressed in terms of binary signals $a_1$ and $a_2$ as $a = (a_1 + a_2) / 2$ where

\begin{equation}
a_1 =
\begin{cases}
  2 & \text{if } a \geq 1 \\
  0 & \text{if } a = 0
\end{cases}
\end{equation}

\begin{equation}
a_2 =
\begin{cases}
  2 & \text{if } a = 2 \\
  0 & \text{if } a \leq 1
\end{cases}
\end{equation}

\begin{figure}[htbp]
	\centering
	\includegraphics[width=0.4\columnwidth]{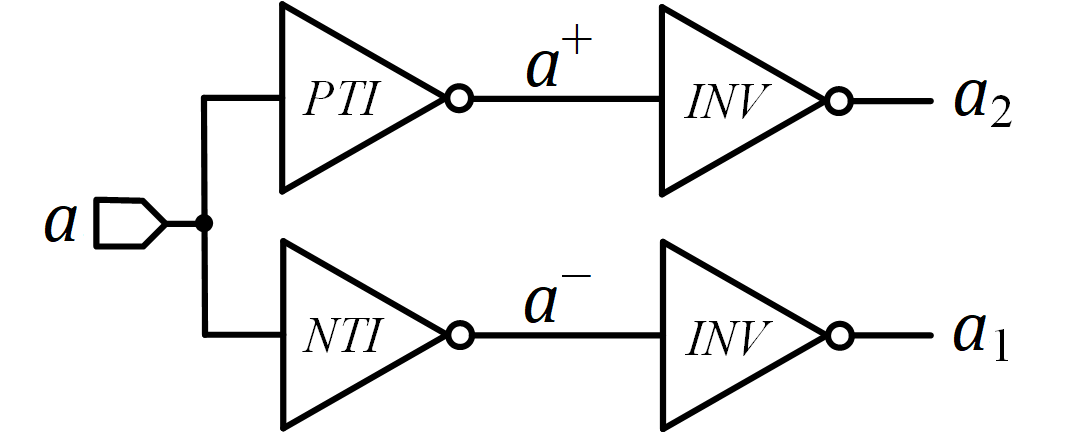}\label{fig:TLGDecode}
	\caption{Decomposing the ternary signal into binary forms $a_1$ and $a_2$ via ternary and binary inverters.}
\end{figure}

We can obtain the decomposed binary signals $a_1$ and $a_2$ by first applying negative ternary inverter (NTI) and positive ternary inverter (PTI) and then binary inverters (INV) as in Fig.~\ref{fig:TLGDecode}. We provide definitions of the negative ternary inverter, $a^-$, and positive ternary inverter, $a^+$, in Eqns. \ref{eqn:NTI} and \ref{eqn:PTI}, respectively. We provide transistor level implementation of these inverters in Appendix~\ref{sec:invTernary}.

\begin{equation}\label{eqn:NTI}
a^-=
\begin{cases}
  2 & \text{if } a= 0 \\
  0 & \text{if } a \neq 0
\end{cases}
\end{equation}

\begin{equation}\label{eqn:PTI}
a^+  =
\begin{cases}
  2 & \text{if } a \neq 2 \\
  0 & \text{if } a = 2
\end{cases}
\end{equation}

\begin{figure}[htbp]
	\centering
	\includegraphics[width=0.65\columnwidth]{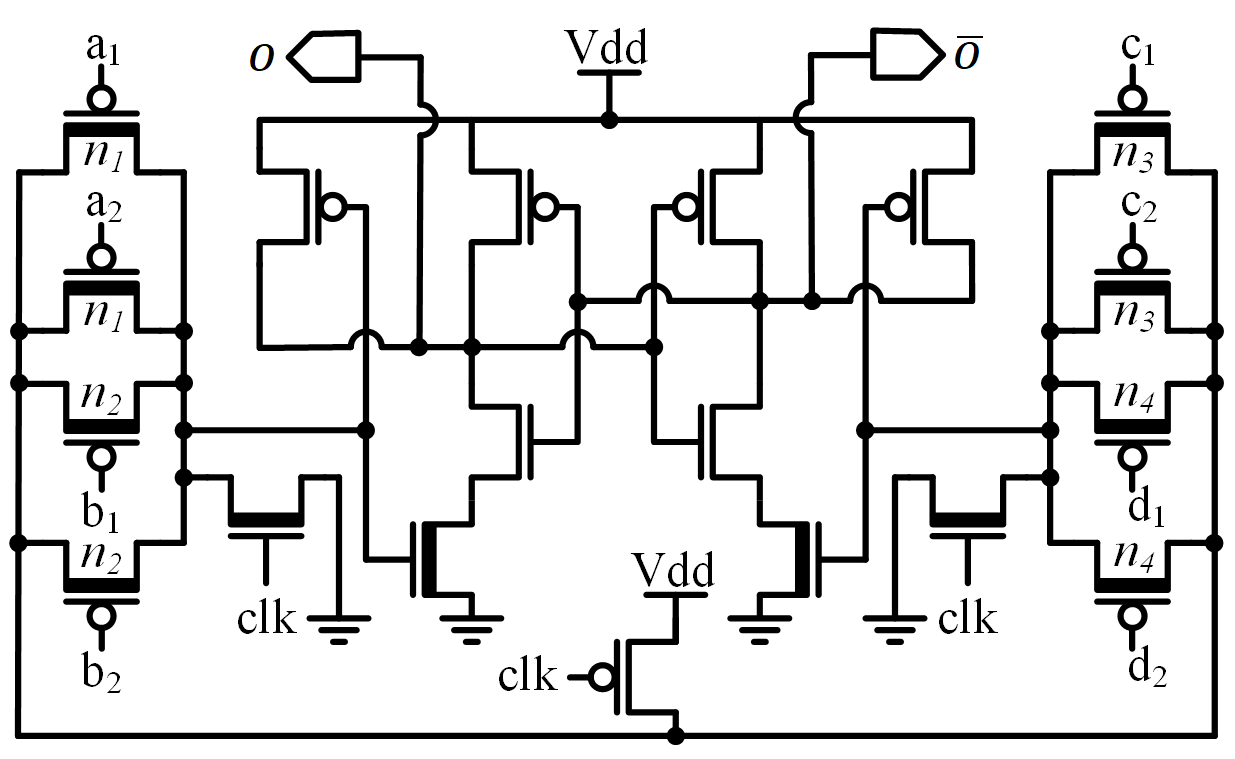}
	\caption{Differential TLG circuit.}\label{fig:transistorDTL}
\end{figure}

We can obtain $b_1$, $b_2$, $c_1$, $c_2$, $d_1$, and $d_2$ signals in the same way from ternary inputs $b$, $c$, and $d$ of the TLG in Fig.~\ref{fig:DTLBlock}. Then, we feed these signals to the differential TLG in Fig.~\ref{fig:transistorDTL}. This way, we can rewrite Eqn.~\ref{eqn:o_and_obar1} as

\begin{equation} \label{eq1}
\begin{split}
o = &\sgn(n_1 \cdot (a_1+a_2)/2 + n_2 \cdot (b_1+b_2)/2 + \varepsilon - n_3 \cdot (c_1+c_2)/2\\
 & - n_4 \cdot (d_1+d_2)/2) + 1
 \end{split}
\end{equation}

The layout in Fig.~\ref{fig:transistorDTL} is a well known differential TLG circuit which works in two phases controlled by a clock \cite{Leshner2010, Kulkarni2016}. When the clock is high, internal nodes connected to the comparator are discharged to ground. When the clock goes low, transistors controlled by the signals $a_i$, $b_i$, $c_i$, and $d_i$ charge internal nodes. The current in-balance between left and right sides of the comparator forces one side to fall to ground. We always inject a small amount of additional current to the right side of the circuit in our implementation. This corresponds to the small positive offset $\varepsilon$ in Eqn.~\ref{eq1}.

To sum up, we implement the proposed ternary TLG by combining the differential TLG circuit in Fig.~\ref{fig:transistorDTL} with the circuit in Fig~\ref{fig:TLGDecode} and its variants for $b$, $c$, and $d$. The circuit in Fig.~\ref{fig:transistorDTL} is level sensitive to clock. We append a binary SR-latch to it to make it edge sensitive. As a result, the proposed TLG based implementation comes with a bonus flip-flop.

\section{Ternary Logic Gates}\label{sec:logicTernary}

Ternary logic gates can be implemented utilizing the proposed ternary TLG. In this section, we introduce ternary TLG based implementations for the ternary inverter and ternary two input AND, OR, and XOR gates. M input gates can be implemented by combining these two input gates hierarchically.

\subsection{Inverter}\label{sec:ternaryNOT}

\begin{figure}[htbp]
	\centering
    \includegraphics[width=0.55\columnwidth]{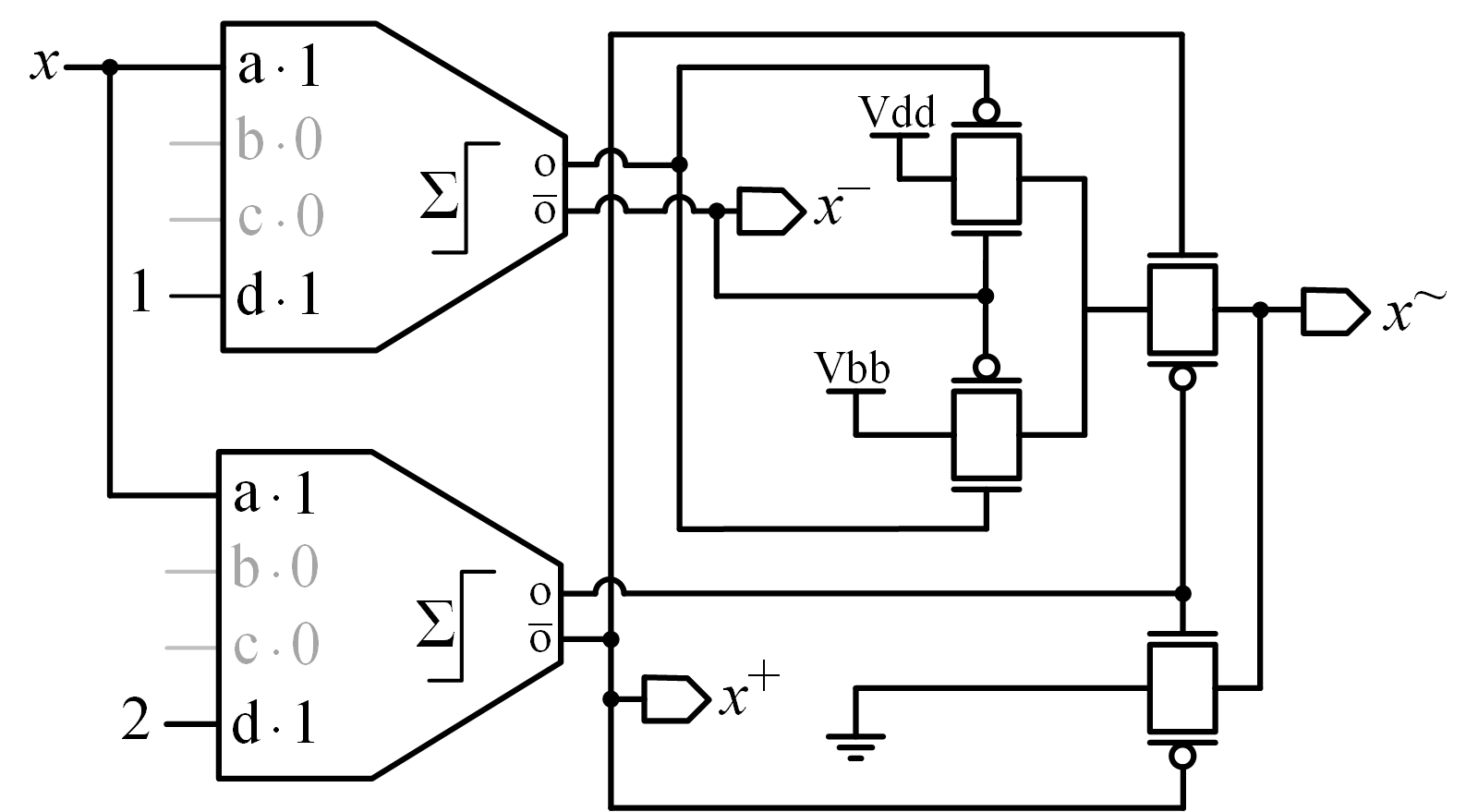}
	\caption{Implementation of STI via the ternary TLG.}\label{fig:tlginv}
\end{figure}

Standard ternary inverter (STI) operation on a trit $x$ is defined as $x^\sim= 2 - x$. We implement this function using the proposed ternary TLG in as Fig~\ref{fig:tlginv}. In this circuit, outputs of TLGs control transmission gates. If the input $x$ is at logic level 2, the output $x^\sim$ is connected to ground which corresponds to logic level 0. When $x$ is at logic level 1, the output is connected to $V_{bb}$ which represents logic level 1. Similarly, the output is shorted to $V_{dd}$ (logic level 2), when $x$ is at logic level 0.

\subsection{AND Gate}\label{sec:ternaryAND}

\begin{figure}[htbp]
	\centering
    \includegraphics[width=0.45\columnwidth]{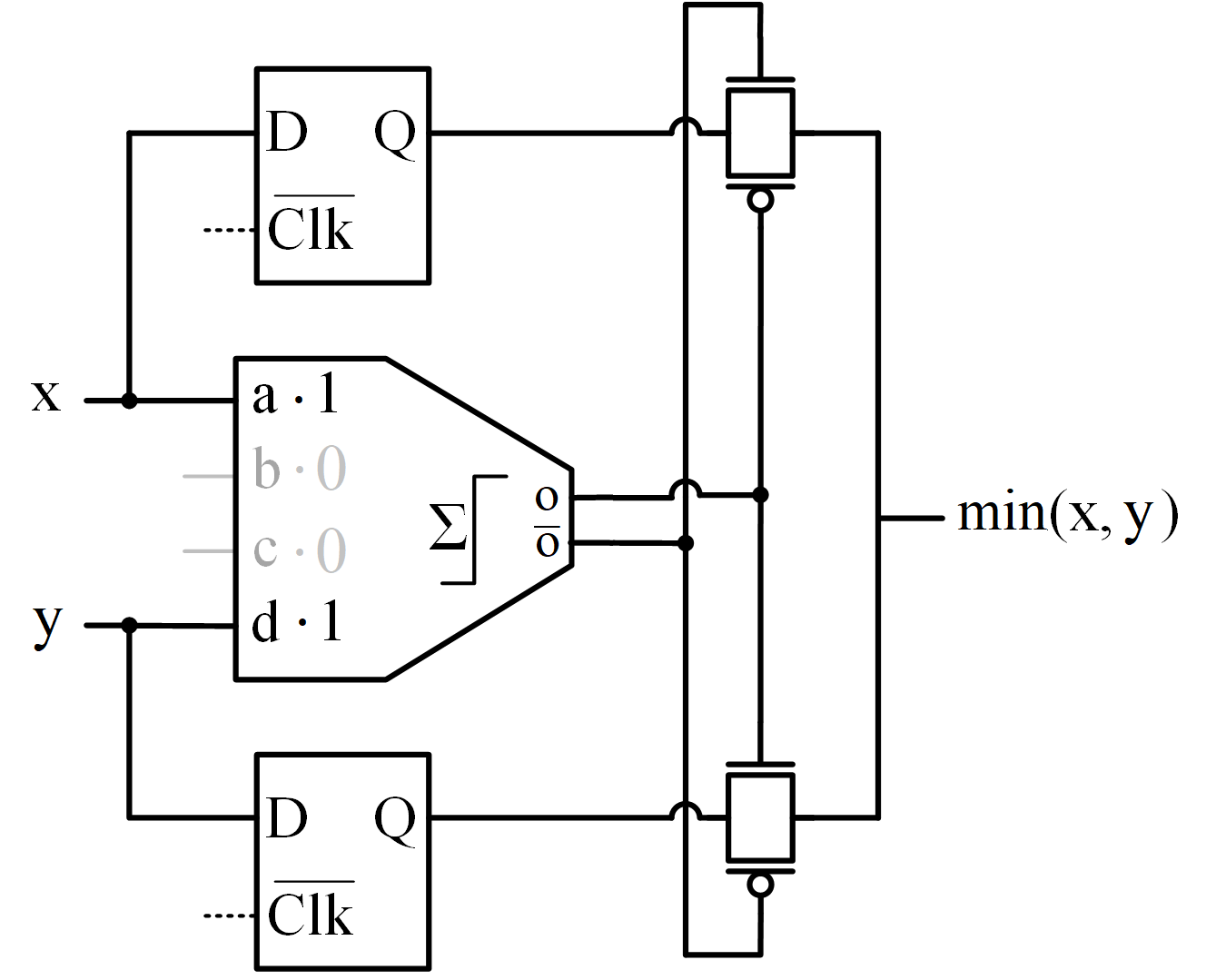}
    \caption{AND gate (min function) implementation for ternary logic utilizing TLG.}\label{fig:min}
\end{figure}

There are different definitions of the ternary AND gate. We pick the $\min(\cdot)$ function as the ternary AND operation in this study. We can re-frame the mathematical representation for this gate as

\begin {equation}
\min (x,y)=\frac{x}{2}(1+\sgn(y-x+\varepsilon))+\frac{y}{2}(1+\sgn(x-y+\varepsilon))
\end {equation}

\noindent where $\sgn(\cdot)$ is the sign function, a small offset $\varepsilon$ is added to prevent the sign function yielding 0. We benefit from the proposed ternary TLG and transmission gates to implement the ternary AND gate as in Fig.~\ref{fig:min}. Here, ternary D-latches are used to synchronize inputs $x$ and $y$ with the clock of the TLG. Implementation of the ternary D-latch is provided in Appendix~\ref{sec:latchTernary}.

\subsection{OR Gate}\label{sec:ternaryOR}

\begin{figure}[htbp]
	\centering
    \includegraphics[width=0.45\columnwidth]{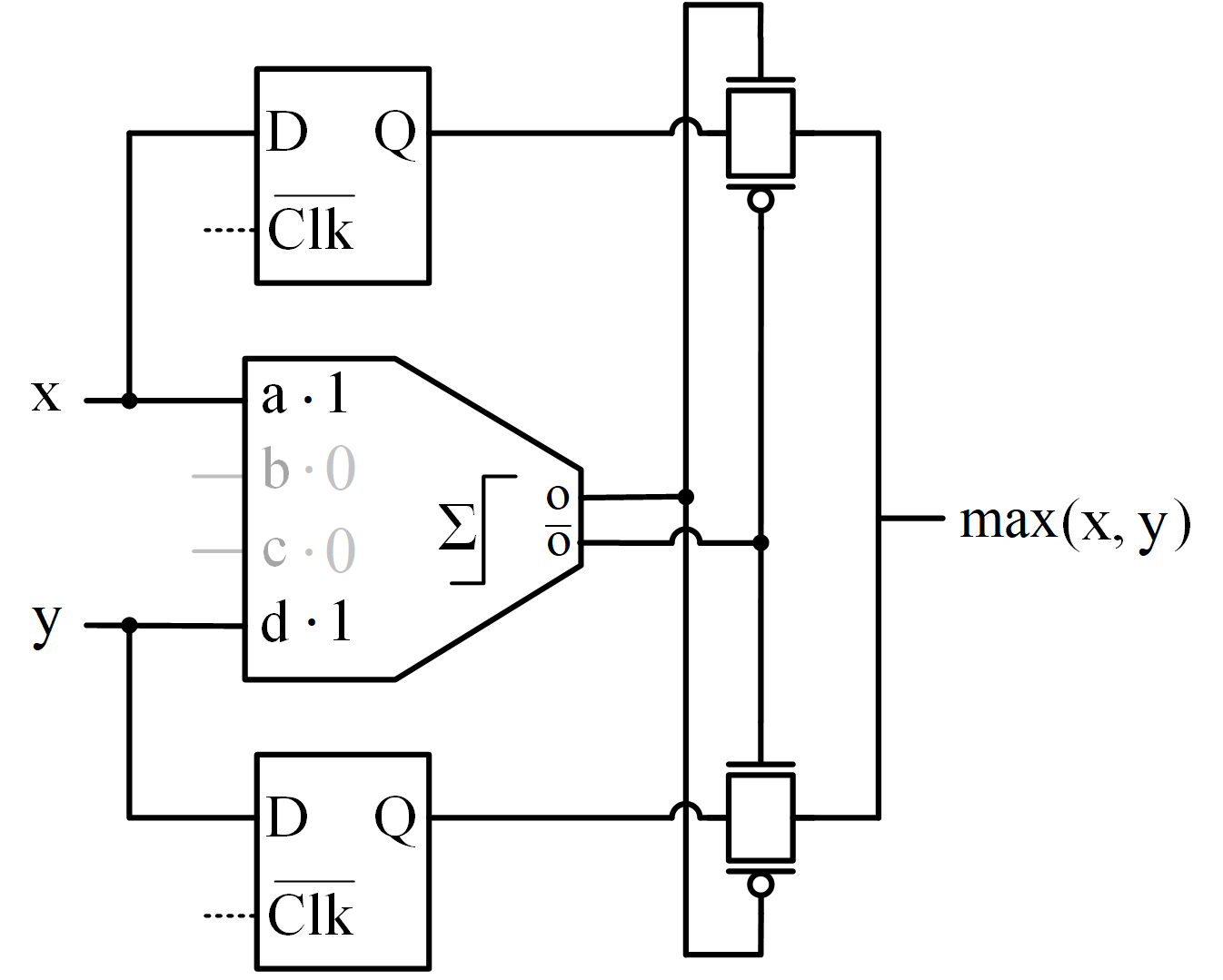}
	\caption{OR gate (max function) implementation for ternary logic utilizing TLG.}\label{fig:max}
\end{figure}

There are different definitions of the ternary OR gate. We use the $\max(\cdot)$ function as the ternary OR operation in this study. We can re-formulate the mathematical representation of this gate as

\begin {equation}
\max(x,y)=\frac{x}{2}(1+\sgn(x-y+\varepsilon))+\frac{y}{2}(1+\sgn(y-x+\varepsilon))
\end {equation}

\noindent where $\sgn(\cdot)$ is the sign function and $\varepsilon$ is a small offset. We benefit from the proposed ternary TLG and transmission gates to implement the $\max (x,y)$ function as in Fig.~\ref{fig:max}. This implementation is similar to the AND gate implementation. However, control signals of the transmission gates are flipped here.

\subsection{XOR Gate}\label{sec:ternaryXOR}

We benefit from the definition in \cite{Murotiya2015} for the XOR gate. Hence, output $s$ of the ternary XOR gate is defined as the modulo-3 sum of two trits $x$ and $y$. The generated carry is ignored. Truth table of the ternary XOR gate is provided in Table~\ref{tbl:xor}.

\begin{table}[htbp]
\centering
\caption{Truth table of the ternary XOR gate.}\label{tbl:xor}
\begin{tabular}{cc|cc|cccccc}
\hline
x & y & s \\
\hline
0 & 0 & 0\\
0 & 1 & 1\\
0 & 2 & 2\\
1 & 0 & 1\\
1 & 1 & 2\\
1 & 2 & 0\\
2 & 0 & 2\\
2 & 1 & 0\\
2 & 2 & 1\\
\hline
\end{tabular}
\end{table}

\begin{figure}[htbp]
	\centering
	\includegraphics[width=0.55\columnwidth]{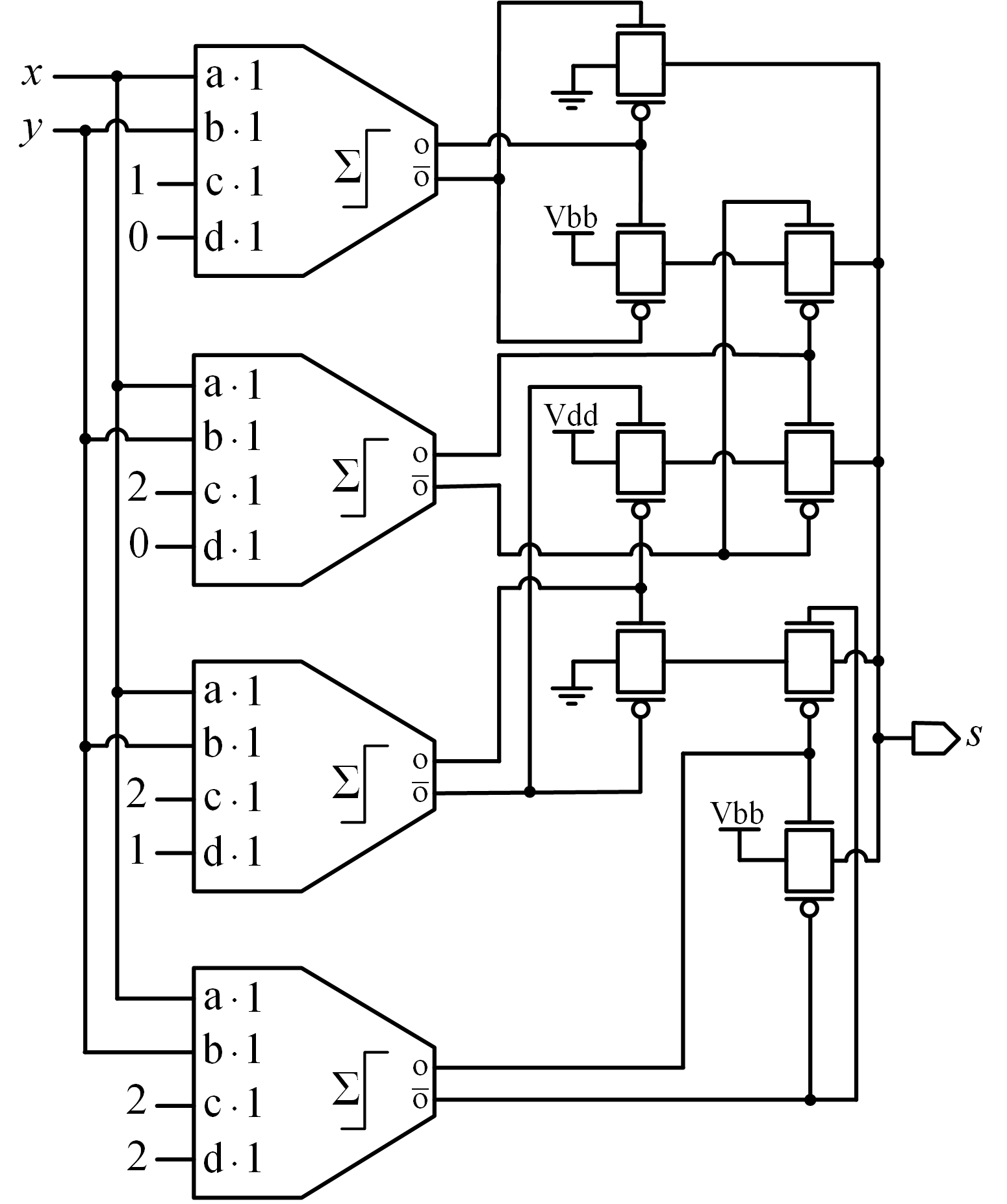}
	\caption{Implementation of the ternary XOR gate via TLG.}\label{fig:tlgxor}
\end{figure}

We implement the ternary XOR gate using the proposed TLG as in Fig.~\ref{fig:tlgxor}. As can be seen in this figure, we short circuit output to ground (logic level 0), or to $V_{bb}$ (logic level 1), or to $V_{dd}$ (logic level 2) via transmission gates. Control signals of transmission gates are generated by ternary TLGs.

\section{Comparing Ternary Numbers}\label{sec:compTernary}

We form a setup to compare two ternary numbers via the proposed ternary TLG. First, we start with the comparison of two trits. Then, we extend this setup to compare two ternary numbers with M digits.

\subsection{Comparison of Two Trits}\label{sec:comp2Trits}

\begin{figure}[htbp]
	\centering
	\includegraphics[width=0.3\columnwidth]{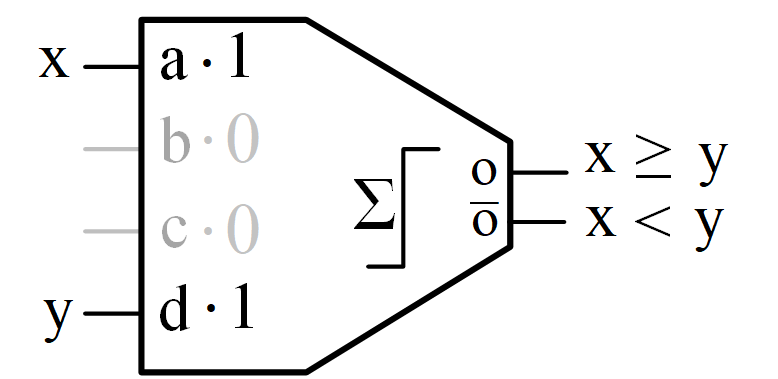}
	\caption{Circuit setup for the comparison of two trits.}\label{fig:comp1}
\end{figure}

We can compare two trits, $x$ and $y$, benefiting from the proposed ternary TLG. The circuit setup for the comparator is as in Fig.~\ref{fig:comp1}. In this figure, $x$ and $y$ are connected to inputs $a$ and $d$ of the TLG, respectively. If the output $o$ is at logic level 2, then we know that $x \geq y$. Otherwise, $x < y$.

If we swap inputs $x$ and $y$ in Fig.~\ref{fig:comp1}, the modified TLG circuit will check $y \geq x$. Thus, using the circuit in the figure and its modified version, we can simultaneously check the conditions $x \geq y$ and $y \geq x$. Moreover, the proposed TLG allows us to use binary logic gates AND ($\wedge$) and OR ($\vee$) since it outputs logic level 0 (false) or 2 (true). This leads to controlling $x == y$ as $(x \geq y)\wedge(y \geq x)$.

\subsection{Comparison of Two Ternary Numbers}\label{sec:comp2Ternary}

The comparator in Fig.~\ref{fig:comp1} can be used to setup a circuit to compare two ternary numbers with M digits. In fact, this comparison is the same as the one used in binary logic. Let's explain it by an example. Assume that we have two ternary numbers $x$ and $y$ with two digits each. Hence, we will have $x=x_2x_1$ and  $y=y_2y_1$ where $x_i$ and $y_i$ for $i=1,2$ are trits. We can check whether $x>y$ by

\begin{equation}
(x_2>y_2) \vee \left( (x_2==y_2) \wedge (x_1>y_1) \right )
\end{equation}

\noindent We can extend this operation to two ternary numbers with M digits in the same way.

\section{Ternary Arithmetic Operations}\label{sec:arthTernary}

We can implement ternary arithmetic operations via the proposed ternary TLG. In this section, we consider the ternary half-adder implementation first. Then, we provide the ternary complement operation to use the same circuitry in implementing the subtraction operation.

\subsection{Ternary Half Adder}

\begin{figure}[htbp]
	\centering
    \includegraphics[width=0.25\columnwidth]{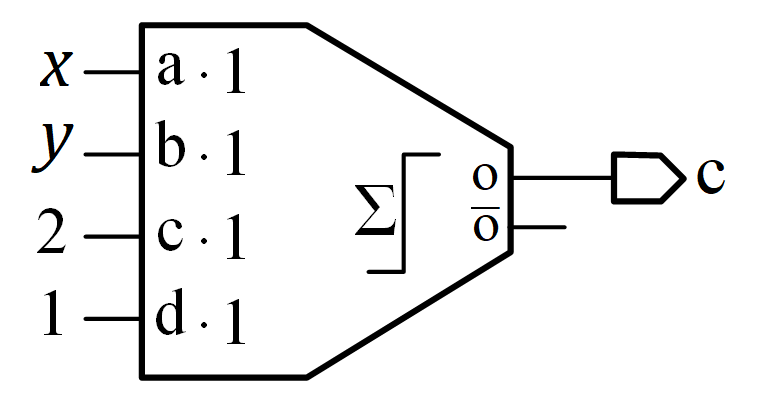}
	\caption{Carry generator circuit of ternary half adder.}\label{fig:carryTLG}
\end{figure}

We propose a ternary half adder (THA) circuit with the proposed ternary TLG in this section. The reader may find truth table of the THA in Appendix~\ref{sec:anyTernary}. We benefit from the ternary XOR gate presented in Section~\ref{sec:ternaryXOR} to calculate the sum trit in the proposed implementation. We use the TLG circuit in Fig.~\ref{fig:carryTLG} to calculate the carry trit.

\subsection{Ternary Half Subtractor}

Complement based subtraction operation used in binary logic applies to ternary (and in general Nary numbers) in the same way. For the ternary case, one's complement in binary logic is replaced by two's complement in ternary logic. Two's complement in binary logic is replaced by three's complement in ternary logic. To perform these operations, we just need ternary inversion. Here, we will use the STI introduced in Section~\ref{sec:ternaryNOT}. Then, we will use THA for complement based subtraction. We should also increment the result. Here, we can use THA as well.


\section{Comparison of Designs}\label{sec:comparison}

We provide comparison of designs in this section. We handle this in two steps. First, we compare the TLG based ternary inverter, comparator, AND, OR, and XOR gates with their standard (CMOS only) implementations. Second, we handle the THA circuit which is frequently used to compare the performance of CNTFET ternary circuits \cite{Sharma2019}. In the same line, we compare our THA with the previously reported THA circuits in this section. We provide the simulation files for all our designs in the GitHub link \url{https://github.com/ahmetunu/Ternary-TLG.git}. Hence, the reader can compare the provided designs with theirs in the future.


\subsection{Inverter, Comparator, AND, OR and XOR Gates}\label{sec:comparison_CMM}

We compare the TLG based ternary inverter (STI), comparator (COMP), AND, OR, and XOR gates with their standard (CMOS only) implementation in terms of simulations in this section. We provide the standard implementation of STI in Appendix~\ref{sec:invTernary}. We use the method in Appendix~\ref{sec:anyTernary} to implement the standard version of COMP, AND, OR, and XOR gates.

We performed all simulations by setting the clock speed to 1~GHz. Besides, all outputs are loaded with 1~fF capacitor. The proposed ternary TLG inherently includes a flip-flop at its output. Therefore, flip-flops are appended to standard implementations to obtain the same functionality and to make a fair comparison. We provide the comparison results between the proposed ternary TLG based designs and their standard implementations in Table~\ref{tab:STDvsTLG}. In this table, STD+FF stands for the standard implementation with flip-flops. TLG stands for the proposed ternary TLG based implementation.

\begin{table}[htbp]
\centering
\caption{Comparison of ternary circuits.}\label{tab:STDvsTLG}
\begin{tabular}{ccrrrrr}
\hline & & \multicolumn{1}{c}{Static P.} & \multicolumn{1}{c}{Power} & \multicolumn{1}{c}{Delay} & \multicolumn{1}{c}{PDP} & \multicolumn{1}{c}{Area} \\
Circuit & Type & $\times10^{-7}$ W & $\times10^{-6}$ W & $\times10^{-12}$ s & $\times10^{-16}$ J & $\times10^{-2}$ $\mathrm{{\mu m}^2}$ \\
\hline
\multirow{2}{*}{\begin{tabular}[c]{@{}c@{}} STI \end{tabular}}    & STD+FF & 19.11 & 10.85 & 270 & 29.30 & 256.77 \\
                                                                                                           & TLG  & \textbf{9.25} & \textbf{7.41} & \textbf{197} & \textbf{14.60} & \textbf{36.54} \\
\hline
\multirow{2}{*}{\begin{tabular}[c]{@{}c@{}} COMP\end{tabular}}  & STD+FF & \textbf{4.63} & 7.51 & 255 & 19.15 & 36.72 \\
                                                                                                             & TLG  & 4.78 & \textbf{3.34} & \textbf{171} & \textbf{5.71} & \textbf{4.05} \\
\hline
\multirow{2}{*}{\begin{tabular}[c]{@{}c@{}} AND \end{tabular}}  & STD+FF & \textbf{21.24} & \textbf{6.93} & 311 & \textbf{21.55} & 246.96 \\
                                                                                                           & TLG  & 21.25 & 9.77 & \textbf{288} & 28.13 & \textbf{238.81} \\
\hline
\multirow{2}{*}{\begin{tabular}[c]{@{}c@{}} OR \end{tabular}}    & STD+FF & \textbf{21.24} & \textbf{7.06} & 309 & \textbf{21.82} & 240.26 \\
                                                                                                           & TLG  & 21.25 & 9.95 & \textbf{304} & 30.25 & \textbf{238.81} \\
\hline
\multirow{2}{*}{\begin{tabular}[c]{@{}c@{}} XOR \end{tabular}}  & STD+FF & \textbf{19.42} & 14.72 & 292 & 42.98 & 259.38 \\
                                                                                                           & TLG  & 24.49 & \textbf{13.65} & \textbf{151} & \textbf{20.61} & \textbf{73.08} \\
\hline
\end{tabular}
\end{table}

As can be seen in Table~\ref{tab:STDvsTLG}, the proposed ternary TLG based designs outperform standard implementations on STI, COMP, and XOR gates. Moreover, proposed ternary TLG based designs for the mentioned circuits offer significant chip area reduction as well as improvement in power consumption and delay. These results are promising and highlight the potential of the proposed ternary TLG based designs. On the other hand, delay and chip area of the TLG based AND and OR gates are similar to their standard implementations. This is mainly due to ternary latches added to these gates for synchronization. If an implementation without those latches can be done, then these gates may also outperform their standard implementation. To note here, ternary TLG based design is completely compatible with standard implementation. Therefore, the designer may use ternary TLG based circuits and standard implementations in the same design.

\subsection{Half Adder}\label{sec:comparison_HA}

We focus on the comparison of proposed TLG based implementation of THA in this section. To the authors knowledge, this is the first study which uses a modern multi-Vt MOSFET technology to implement ternary logic. Therefore, multi-Vt MOSFET ternary designs are not available for comparison. Hence, we compare our design with the CNTFET based designs. To do so, we benefit from the results provided in \cite{Sharma2019}. We also compare TLG based design with its standard implementation given in Appendix~\ref{sec:anyTernary}.

During comparison, all simulations are performed at~500 MHz and all outputs are loaded with 1~fF capacitor to be consistent with the setup in \cite{Sharma2019}. We provide the comparison results in Table~\ref{tab:tha}. In this table, we compare power consumption and delay of the proposed ternary TLG based THA implementation with the standard implementation (STD), standard implementation having flip-flops (STD+FF), and the ones in literature \cite{Bastani2017, Lin2009, Srinivasu2016, Sahoo2017, Sharma2019}. Unfortunately, CNTFET based designs are abstract and their chip area is not available. Therefore, we could not perform the area comparison for the THA.

\begin{table}[htbp]
\centering
\caption{Comparison of THA circuits.}\label{tab:tha}
\begin{tabular}{cccrrrr}
\hline
\multicolumn{1}{c}{Ternary} & \multicolumn{1}{c}{Device} & \multicolumn{1}{c}{Flip} & \multicolumn{1}{c}{Static P.} & \multicolumn{1}{c}{Power} & \multicolumn{1}{c}{Delay} & \multicolumn{1}{c}{PDP} \\
\multicolumn{1}{c}{Half Adder} & \multicolumn{1}{c}{Type} & \multicolumn{1}{c}{Flop} & \multicolumn{1}{l}{$\times10^{-7}$W} & \multicolumn{1}{c}{$\mu$W} & \multicolumn{1}{l}{$\times10^{-12}$s} & \multicolumn{1}{l}{$\times10^{-16}$J}  \\
\hline
Bastani~\emph{et al.}~\cite{Bastani2017} & CNTFET & No & 52.5 & 2.74 & 69.8 & 1.91 \\
Lin~\emph{et al.}~\cite{Lin2009} & CNTFET & No & 76.8 & 4.00 & 65.9 & 2.64 \\
Srinivasu~\emph{et al.}~\cite{Srinivasu2016} & CNTFET & No & 183.6 & 95.88 & 72.2 & 69.24 \\
Sahoo~\emph{et al.}~\cite{Sahoo2017} & CNTFET & No & 16.2 & \textbf{1.15} & 50.1 & \textbf{0.57} \\
Sharma~\emph{et al.}~\cite{Sharma2019} & CNTFET & No & 6.7 & 1.43 & \textbf{39.7} & \textbf{0.57} \\
\hline
STD & MOSFET & No & \textbf{4.9} & 1.19 & 184.2 & 4.40 \\
STD+FF & MOSFET & \textbf{Yes} & 19.5 & 8.56 & 243.6 & 20.85 \\
TLG (Proposed) & MOSFET & \textbf{Yes} & 24.5 & 7.77 & 185.9 & 14.44 \\
\hline
\end{tabular}
\end{table}

We can analyze the results in Table~\ref{tab:tha} in detail. Beforehand, we should mention that none of the CNTFET based designs have a flip-flop at their output. Therefore, the comparison is significantly biased and favors CNTFET implementations. To reduce this bias, we included a standard (CMOS only) implementation (STD) without flip-flops. Comparing the standard implementation and CNTFET designs, we observe that power consumption of these are similar. Hence, we expect to see similar power figures for the ternary TLG based design and CNTFET designs if flip-flops are added to CNTFET implementations.

Comparing the delay of CNTFET based designs with the standard implementation in Table~\ref{tab:tha}, we may conclude that CNTFET technology offers less delay. Although CNTFET technology seems promising, large scale CNTFET designs do not seem to be achievable in the near future. The largest CNTFET design is presented in \cite{Hills2019} and it only contains 15.000 CNTFETs. Moreover, integrating CNTFETs with different threshold voltages is an additional challenge. On the other hand, designs implemented with recent CMOS technology nodes contain several billions of transistors. Moreover, these technology nodes readily support multi-threshold designs. Thus, in contrary to CNTFET based designs, our designs based on the proposed TLG can be realized physically. Therefore, we have a significant advantage compared to CNTFET based designs. As can be seen in Table~\ref{tab:tha}, our ternary TLG based THA design decreases power consumption and delay compared to standard implementation with the same functionality (STD+FF).

\section{Final Comments}\label{section:final_comments}

In this study, we propose the first implementation of a ternary TLG circuit which is designed using a multi-Vt CMOS process. Using this gate, we implemented the ternary inverter, comparator, AND, OR, and XOR gates, arithmetic operators for ternary addition and subtraction. These modules can be taken as subparts of an ALU working on ternary representations. We compared the proposed TLG based implementations with their standard (CMOS only) versions and the ones in literature. We picked the design metrics as power consumption, delay, and chip area in comparison. The proposed TLG based inverter, comparator, XOR gates and THA outperformed their standard implementations. Moreover, TLG based design come with a bonus flip-flop. This makes it very suitable for RTL designs, where flip-flops are frequently required. However, the proposed ternary TLG based implementation could not yield a significant improvement for the ternary AND and OR gates due to additional latches. Hence, a design methodology is needed to decide when to use the proposed ternary TLG. Utilizing such a design methodology, appropriate parts of a ternary CMOS may be replaced with the TLG. This may lead to improving the delay, power consumption, and area of ternary circuits.

\bibliography{ternary_NN}

\newpage

\begin{appendices}

\section{Standard Implementation of Ternary Functions}

We provide the standard (CMOS only) implementations of ternary functions in this section. Our aim here is twofold. First, some of these designs have been used in implementation of the proposed ternary TLG. Second, we compare these standard designs with the ones proposed in this study.

\subsection{Ternary Inverters}\label{sec:invTernary}

\begin{figure}[htbp]
	\centering
	\includegraphics[width=\columnwidth]{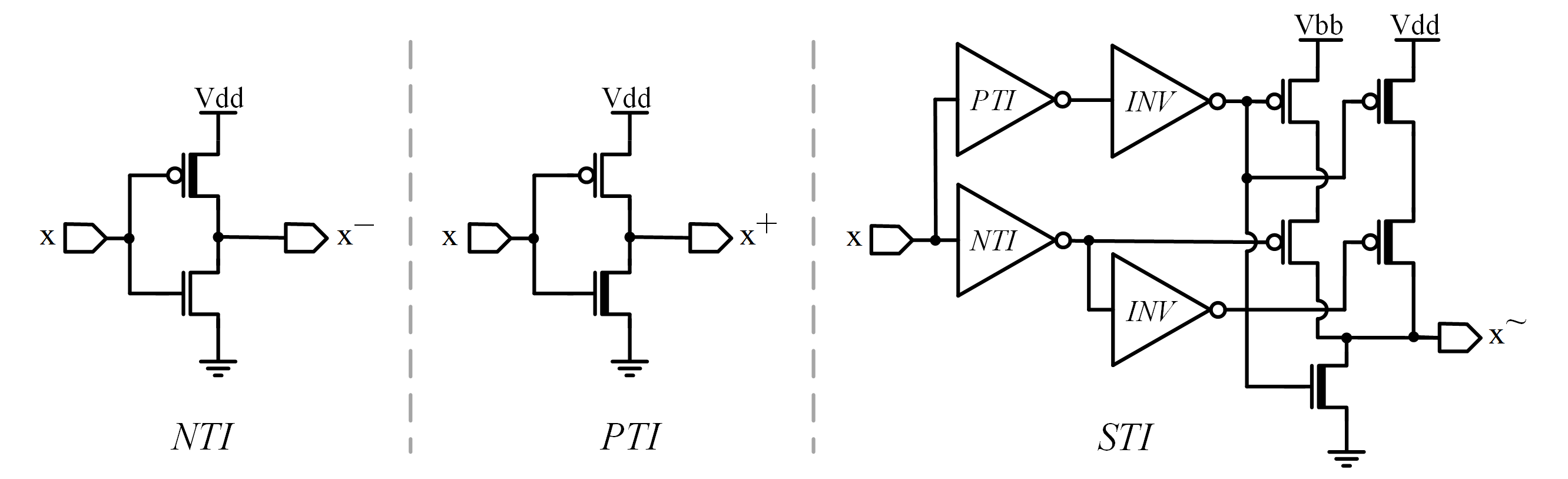}
	\caption{Transistor level implementation of ternary inverters NTI, PTI, and STI.}\label{fig:invsTernary}
\end{figure}

Transistor level implementation of the ternary inverters NTI, PTI, and STI are given in Fig.~\ref{fig:invsTernary}. In these designs, low and high threshold devices are used. The INV gate in the figure represents the binary NOT operation.

\subsection{General Ternary Function}\label{sec:anyTernary}

A ternary function, such as THA, can be implemented via CMOS logic. Therefore, we should check the truth table of ternary inverters and their derivatives as given in Table~\ref{tbl:inv}.

\begin{table}[htbp]
\centering
\caption{Truth table of ternary inverters, NTI, STI, and PTI.} \label{tbl:inv}
\begin{tabular}{ccccccc}
\hline
x & $x^-$ & $x^\sim$ & $x^+$ & $(x^-)^\sim$ & $(x^+)^\sim$ & $x^+\wedge (x^-)^\sim$\\
\hline
0 & \textbf{2} & 2 & 2 & 0 & \textbf{0} &\textbf{0}\\
1 & \textbf{0} & 1 & 2 & 2 &\textbf{0} & \textbf{2}\\
2 & \textbf{0} & 0 & 0 & 2 &\textbf{2} & \textbf{0}\\
\hline
\end{tabular}
\end{table}

As can be seen in Table~\ref{tbl:inv}, $x^-$ is active iff $x=0$. $(x^+)^\sim$ is active iff $x=2$. $x^+\wedge (x^+)^\sim$ is active iff $x=1$. Therefore, these three representations are sufficient to decode $x$ as

\begin{equation}\label{eqn:decode}
x =
\begin{cases}
  0 & \text{if } x^- \\
  1 & \text{if } x^+\wedge (x^+)^\sim \\
  2 & \text{if } (x^+)^\sim
\end{cases}
\end{equation}

\begin{figure}[htbp]
	\centering
	\includegraphics[width=0.4\columnwidth]{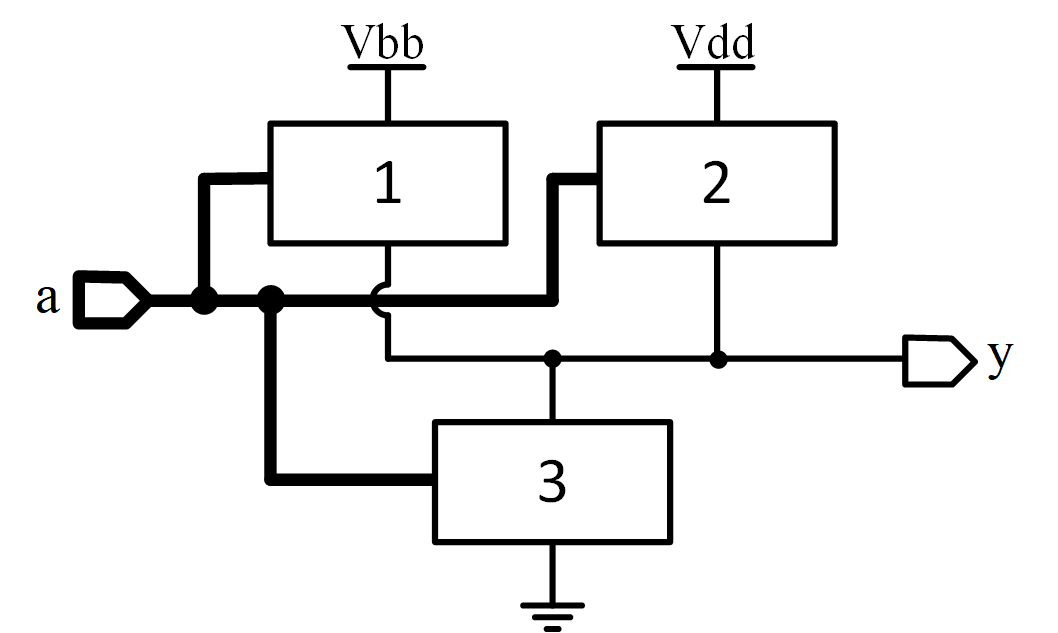}
	\caption{General structure to implement a ternary function.}\label{fig:ternaryNetwork1}
\end{figure}

Based on the decoding scheme in Eqn.~\ref{eqn:decode}, a general structure to implement a ternary function can be formalized as in Fig.~\ref{fig:ternaryNetwork1}. There are three networks in this figure labeled as 1, 2, and 3. The first network is activated when $y(a)=1$. In this case, this network pulls the output $y$ to $V_{bb}$. The second network is activated when $y(a)=2$. In this case, the network pulls the output $y$ to $V_{dd}$. The third network is activated when $y(a)=0$. In this case, the third network pulls the output $y$ to ground.

As an example, we can implement the THA following the presented methodology. Truth table of the THA and decoded signals are given in Table~\ref{tbl:tha}. In this table, $c$ and $s$ stand for the carry and sum trits, respectively.

\begin{table}[htbp]
\centering
\caption{Truth table of the THA.}
\begin{tabular}{cc|cc|cccccc}
\hline
x & y & c & s & $x^-$ & $y^-$ & $(x^+)^\sim$ & $(y^+)^\sim$ & $x^+\wedge(x^-)^\sim$ & $y^+\wedge(y^-)^\sim$\\
\hline
0 & 0 & 0 & 0 & 2 & 2 & 0 & 0 & 0 & 0\\
0 & 1 & 0 & 1 & 2 & 0 & 0 & 0 & 0 & 2\\
0 & 2 & 0 & 2 & 2 & 0 & 0 & 2 & 0 & 0\\
1 & 0 & 0 & 1 & 0 & 2 & 0 & 0 & 2 & 0\\
1 & 1 & 0 & 2 & 0 & 0 & 0 & 0 & 2 & 2\\
1 & 2 & 1 & 0 & 0 & 0 & 0 & 2 & 2 & 0\\
2 & 0 & 0 & 2 & 0 & 2 & 2 & 0 & 0 & 0\\
2 & 1 & 1 & 0 & 0 & 0 & 2 & 0 & 0 & 2\\
2 & 2 & 1 & 1 & 0 & 0 & 2 & 2 & 0 & 0\\
\hline
\end{tabular}
\label{tbl:tha}
\end{table}

\begin{figure}[htbp]
	\centering
    \includegraphics[width=0.7\columnwidth]{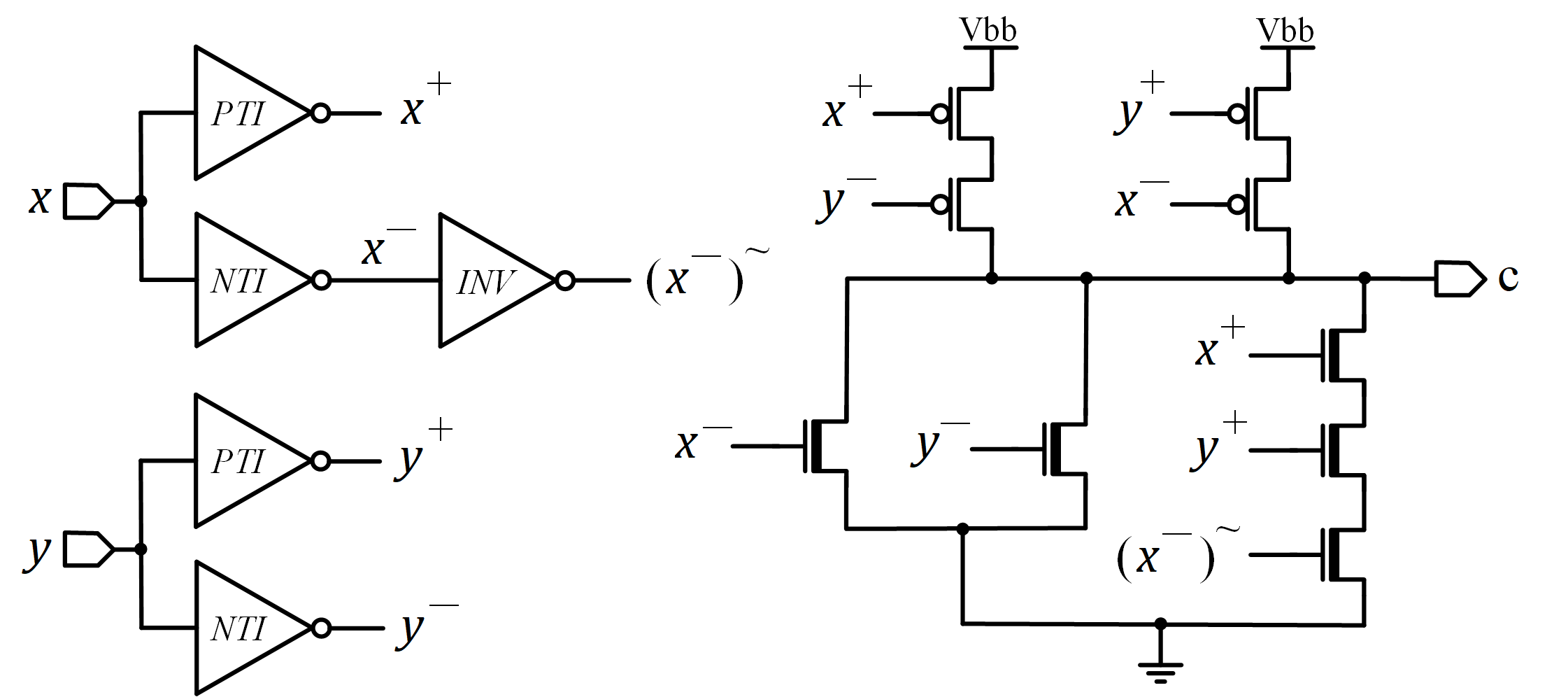}
	\caption{Implementation of the carry generator circuit.}\label{fig:carrySTD}
\end{figure}

As can be seen in Table~\ref{tbl:tha}, the $c$ trit is 0 when $x^- \vee y^- \vee (x^+\wedge(x^-)^{\sim}\wedge y^+)$ is true. Based on this, we can implement the pull down network (network 3). Similarly, we can design the remaining networks (network 2 and 3) for the carry output $c$ of THA. The circuit to generate the carry $c$ is shown in Fig.~\ref{fig:carrySTD}. Following a similar procedure, we can implement the circuit for the sum output $s$ of the THA.

\subsection{Ternary D-Latch}\label{sec:latchTernary}

\begin{figure}[htbp]
	\centering
	\includegraphics[width=0.65\columnwidth]{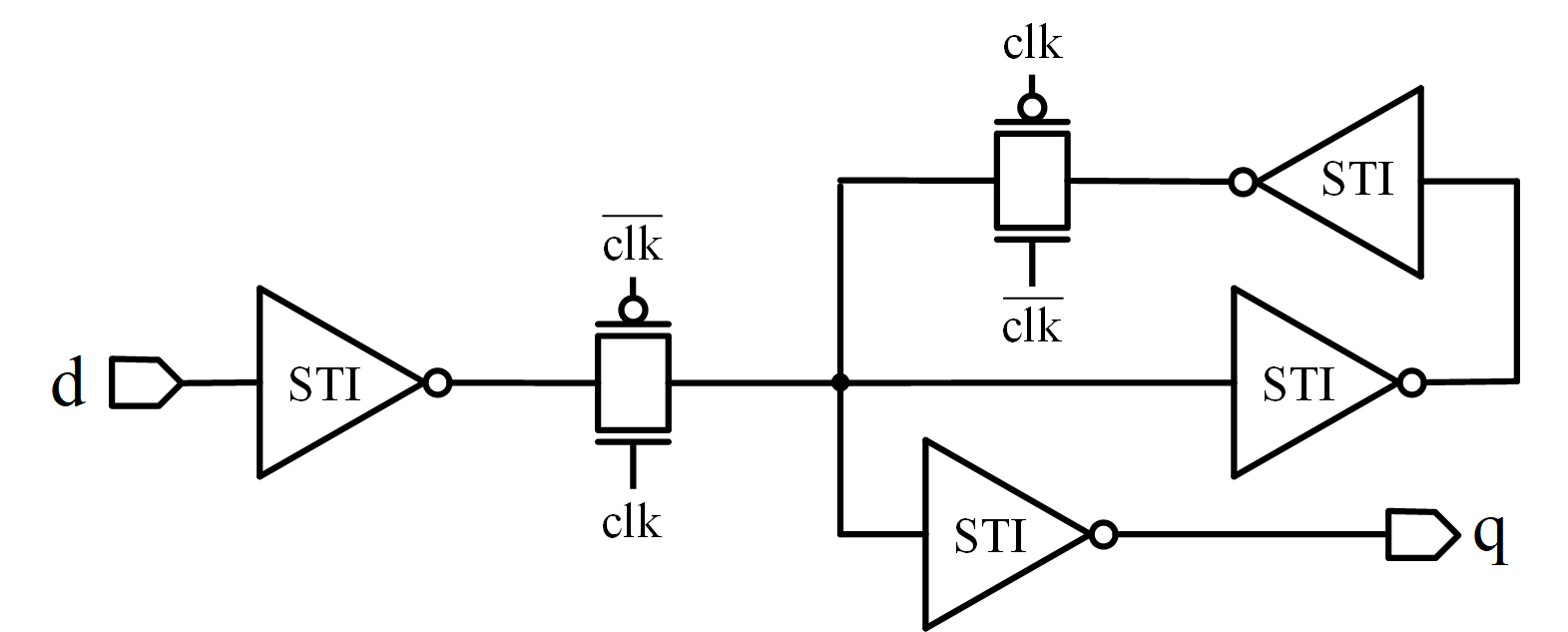}
	\caption{Implementation of the ternary D-latch.}\label{fig:latchTernary}
\end{figure}

Ternary D-latch may be implemented by using two cross coupled STI gates given in Appendix~\ref{sec:invTernary}. Circuit implementation of the ternary D-latch is as in Fig.~\ref{fig:latchTernary}. Working principles of this latch can be explained based on the circuit in this figure as follows. When the clock signal is high, the transmission gate on the left side of the circuit conducts and the transmission gate on the right side turns off. Hence, the feedback loop is cut. When the clock signal is low, connection to the input is cut and the feedback is activated. Thus, two cross coupled STI gates can hold trit values 0, 1, or 2. By concatenating two ternary latches, a ternary D flip-flop can be implemented.

\end{appendices}
\end{document}